\documentclass[twocolumn,journal]{IEEEtran}
\usepackage[T1]{fontenc}
\usepackage[latin9]{inputenc}
\usepackage{textcomp}
\usepackage{color}
\usepackage{amsbsy}
\usepackage{amstext}
\usepackage{graphicx}
\usepackage{fancyhdr}
 
\usepackage[unicode=true,
 bookmarks=true,bookmarksnumbered=true,bookmarksopen=true,bookmarksopenlevel=1,
 breaklinks=false,pdfborder={0 0 0},pdfborderstyle={},backref=false,colorlinks=false]
 {hyperref}
\hypersetup{pdftitle={Your Title},
 pdfauthor={Your Name},
 pdfpagelayout=OneColumn, pdfnewwindow=true, pdfstartview=XYZ, plainpages=false}
\usepackage{breakurl}

\makeatletter
\let\oldforeign@language\foreign@language
\DeclareRobustCommand{\foreign@language}[1]{%
  \lowercase{\oldforeign@language{#1}}}

\usepackage[caption=false,font=footnotesize]{subfig}

\makeatother

\begin{document}

\title{Distributed Degenerate Band Edge Oscillator}

\author{Ahmed F. Abdelshafy, Dmitry Oshmarin, Mohamed A. K. Othman,~Michael M. Green,
and Filippo Capolino}%,~\IEEEmembership{Fellow,~IEEE}}

%\markboth{Abdelshafy \MakeLowercase{\emph{et al.}}: Distributed Degenerate
%Band Edge Oscillator}{}

\maketitle
\makeatletter\renewcommand*\@makefnmark{} \footnotetext{This material is based upon work supported by the Air Force Office of Scientific Research award number FA9550-18-1-0355, and by the National Science Foundation under award NSF ECCS-1711975. The authors are thankful to DS SIMULIA for providing CST Studio Suite that was instrumental in this study. 
} \makeatother

\makeatletter\renewcommand*\@makefnmark{} \footnotetext{The authors are with the Department of Electrical Engineering and Computer Science, University of California, Irvine, CA 92697 USA. (e-mail: abdelsha@uci.edu, doshmari@uci.edu, mothman@uci.edu, mgreen@uci.edu, f.capolino@uci.edu).} \makeatother

%  AF Arxiv
\thispagestyle{fancy}
% Added AF Arxiv

\begin{abstract}
We propose a new class of oscillators by engineering the dispersion
of two-coupled periodic waveguides to exhibit a degenerate band edge
(DBE). The DBE is an exceptional point of degeneracy (EPD) of order
four, i.e., representing the coalescence of four eigenmodes of a waveguide
system without loss and gain. We present a distributed DBE oscillator
realized in periodic coupled transmission lines with a unique mode
selection scheme that leads to a stable single-frequency oscillation,
even in the presence of load variation. The DBE oscillator potentially
leads to a boost of the efficiency and performance of RF sources,
due to the unique features associated to the EPD concept. This class
of oscillators is promising for improving discrete-distributed coherent
sources and can be extended to radiating structures to achieve a new
class of active integrated antenna arrays.
\end{abstract}

\begin{IEEEkeywords}
Coupled transmission line, dispersion engineering, degenerate band
edge, RF oscillator, ladder oscillator
\end{IEEEkeywords}

\section{Introduction}

\IEEEPARstart{O}{scillators} are one of the fundamental components
that exist in any radio frequency (RF) system. Typically, an RF oscillator
is an amplifier with positive feedback mechanism utilizing a gain
device with a selective resonance circuit that generates a single
tone frequency. The negative conductance, i.e., the gain component,
required for positive feedback can be obtained using transistors as
in cross-coupled transistor pair \cite{razavi_design_2012}, or by
circuit topologies such as Pierce, Colpitts, and Gunn diode waveguide
oscillators \cite{pierce_piezoelectric_1923,collin_foundations_2007}.
In pursuance of improving the performance of RF and microwave sources,
many research avenues are currently being investigated \cite{endo_mode_1976,hajimiri_design_1999,wu_silicon-based_2001,razavi_rf_1998,rohde_design_2005}.
The focus of this paper is on a new class of oscillators whose architecture
features i) a cavity made of a periodic coupled-mode waveguide utilizing
a special kind of degeneracy in its dispersion diagram, used as the
passive circuit responsible for frequency selection, and ii) a set
of distributed active devices incorporated in the cavity that provide
the sufficient negative conductance to compensate the losses and thus
to start the oscillation.

Generally, electromagnetic guiding structures or resonators are characterized
by evolution equations that describe the spatial evolution of their
eigenstates (eigenvalues and eigenvectors). We are interested in a
very special degeneracy condition that occurs when two or more of
these eigenstates coalesce into a single degenerate eigenmode at a
certain point in the parameter space \cite{figotin_gigantic_2005,othman_theory_2017,ruter_observation_2010,guo_observation_2009-1}.
Such points are called exceptional points of degeneracy (EPD), and
the order of the EPD is determined by the number of eigenmodes that
\textcolor{black}{coalesce} at this point. The dispersion relation of
eigenmodes in such a guiding structure that exhibits an EPD of order
$n$ has the unique behavior in which $(\omega-\omega_{e})\propto(\beta-\beta_{e})^{n}$
in the vicinity of EPD, where $\omega$ and $\beta$ are the angular
frequency and the propagation constant, respectively, at the EPD they
are denoted by the subscript $e$. In lossless waveguides, as in this
paper, this unique degenerate dispersion behavior is accompanied by
supreme characteristics including the vanishing of the group velocity
\cite{figotin_frozen_2006,gutman_slow_2012} as well as the dramatic
improvement in the local density of states \cite{othman_giant_2016}
resulting in a robust increase in the loaded quality factor of the
structure.

In this paper we focus on the degenerate band edge (DBE), which is
a fourth-order EPD manifested at the band edge of a lossless structure.
\textcolor{black}{At the DBE four eigenstates coalesce into a degenerate
one and the dispersion relation near the DBE is characterized by $(1-\omega/\omega_{e})\approx\eta_{e}(1-k/k_{e})^{4}$
where $\eta_{e}$ is a dimensionless flatness parameter. A finite-length
waveguide made of a finite number $N$ of unit cells forms a Fabry-Perot
cavity (FPC) with a resonance very near the DBE frequency, see details
in \cite{figotin_gigantic_2005,othman_giant_2016,abdelshafy_exceptional_2019},
and the field creates a standing wave. The resonance closest to the
DBE frequency occurs at $\omega_{r,e}$ and for large $N$ it is approximated
by $\omega_{r,e}/\omega_{e}\approx1-\eta_{e}/N^{4}$. Associated to
such DBE resonance, the cavity experiences a field enhancement at
its geometrical center which leads to an enhancement in the $Q$-factor
and less sensitivity to loads }\cite{othman_theory_2017}. The motivation
behind introducing a DBE-based oscillator is based on previous work
related to high power microwave devices; it has shown an enhancement
of gain in electron beam devices based on waveguide with a DBE \cite{othman_giant_2016,abdelshafy_electron-beam-driven_2018}
and demonstrated a low starting (threshold) current and a unique threshold
scaling with length compared to conventional backward wave oscillators
\cite{othman_low_2016,abdelshafy_electron-beam-driven_2018}.

In this paper, we present an example of a DBE oscillator based on
two periodic coupled transmission lines (CTLs) as in one of the configurations
proposed in \cite{abdelshafy_exceptional_2019} and shown in Fig.1(a).
The DBE concept in CTLs was introduced in \cite{Volakis_2009} for
antenna minimization application. We first show the dispersion of
the coupled waveguide where the DBE occurs at several points in the
shown frequency range and beyond. Then, we consider a cavity made
of a finite-length CTL where discrete distributed gain is introduced
leading to a single-frequency of oscillation. We show the robustness
of this new class of oscillators against load variation.

\begin{figure}[t]
\centering{}\centering\includegraphics[width=0.85\columnwidth]{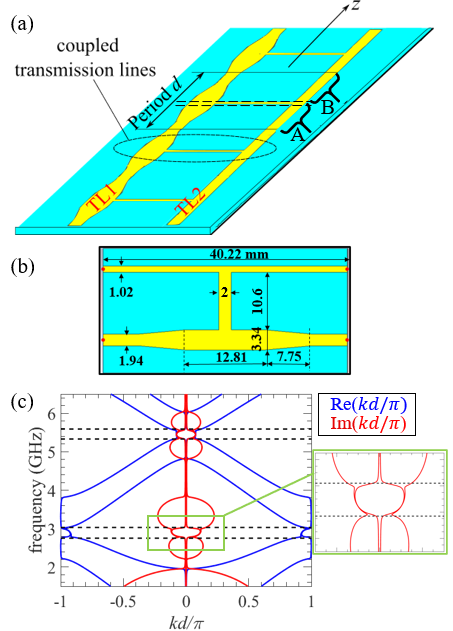}\caption{\label{fig:Example-geometry}(a) Schematic of two coupled microstrip
lines on a grounded dielectric substrate that support a fourth order
EPD (the DBE) visible in the ($k$-$\omega$) dispersion diagram.
(b) Microstrip unit cell of the periodic waveguide that exhibits the
DBE. (c) Real and imaginary parts of the wavenumbers of the four guided
Floquet-Bloch modes obtained using the full-wave finite element method
accounting for radiation, ohmic and dielectric losses.}
\end{figure}

The passive waveguide (before the introduction of distributed gain)
consists of two coupled microstrips over a grounded dielectric substrate
engineered to exhibit a DBE, as shown in Fig. \ref{fig:Example-geometry}(a).
In such a waveguide there are two modal fields that can propagate
along the $+z$-direction, and two in the opposite one.\textcolor{black}{{}
Proper coupling among the four modal fields is required to exhibit
a DBE. The realization of proper coupling is achieved either with
proximity fields (inductive/capacitive coupling) or with a physical
electric connection as in this paper. For more detailed on the engineering
of the proper coupling with an approximate LC model see \cite{abdelshafy_exceptional_2019}
and references therein. The \textquotedblleft corrugation\textquotedblright{}
in TL1 (shown in Fig. 1(a)) is introduced to add another degree of
freedom to our design space parameters. }

The formulation that describes the field evolution using a coupled
transmission line approach is found in \cite{othman_theory_2017,abdelshafy_exceptional_2019},
assuming a time harmonic evolution $e^{j\omega t}$. It is convenient
to define a four-dimensional state vector $\boldsymbol{\Psi}(z)=[\begin{array}{cc}
\mathbf{V}(z), & \mathbf{I}(z)\end{array}]^{T}$, which comprises the voltages $\mathbf{V}(z)=[\begin{array}{cc}
V_{1}(z), & V_{2}(z)\end{array}]^{T}$ and currents $\mathbf{I}(z)=[\begin{array}{cc}
I_{1}(z), & I_{2}(z)\end{array}]^{T}$ in the two lines. The first order differential equations that describe
the spatial evolution of the state vector in a uniform segment of
the CTL are written as \cite{othman_theory_2017}
\begin{equation}
\partial_{z}\boldsymbol{\Psi}(z)=-j\underline{\mathbf{M}}\boldsymbol{\Psi}(z)\label{eq:state_vector_evolution_eqn}
\end{equation}
where
\begin{equation}
\underline{\mathbf{M}}=\left[\begin{array}{cc}
\mathbf{0} & -j\mathbf{Z}\\
-j\mathbf{Y} & \mathbf{0}
\end{array}\right]\label{eq:System_matrix}
\end{equation}
is the 4\texttimes 4 CTL system matrix, and $\mathbf{Z}$ and $\mathbf{Y}$
are the distributed series impedance and shunt admittance 2\texttimes 2
matrices describing the per unit length parameters of the two CTLs
\cite{miano_transmission_2001,othman_theory_2017}. The solution of
\ref{eq:state_vector_evolution_eqn}, assuming a certain boundary
condition at $z=z_{0}$, is found as $\boldsymbol{\Psi}(z_{1})=\mathbf{\underline{T}}(z_{1},z_{0})\boldsymbol{\Psi}(z_{0})$
where $\mathbf{\underline{T}}(z_{1},z_{0})$ is the transfer matrix
given by $\mathbf{\underline{T}}(z_{1},z_{0})=\textrm{exp}(-j(z_{1}-z_{0})\underline{\mathbf{M}})$.

The periodic structure depicted in Fig. \ref{fig:Example-geometry}(a)
has a unit cell composed of two uniform segments A, B and incorporates
an additional coupling matrix due to the coupling microstrip. The
transfer matrix of a unit cell is expressed as the product of the
individual segments' transfer matrices as $\mathbf{\underline{T}}_{\textrm{U}}=\mathbf{\underline{T}}_{\textrm{A}}\mathbf{\underline{T}}_{\textrm{C}}\mathbf{\underline{T}}_{\textrm{B}}$,
where $\mathbf{\underline{T}}_{\textrm{A}}\textrm{ and }\mathbf{\underline{T}}_{\textrm{B}}$
are the transfer matrices of segments A and B, and $\mathbf{\underline{T}}_{\textrm{C}}$
is the coupling matrix representing the physical connection via a
microstrip between the two lines.

\section{Distributed Degenerate Band Edge Oscillator}

The CTL shown in Fig. \ref{fig:Example-geometry} is designed on a
grounded substrate with relative dielectric constant $\epsilon_{r}=$~2.2,
height of 0.508 mm, and with a loss tangent of 0.002. The complex
wavenumber-frequency dispersion diagram shown in Fig. \ref{fig:Example-geometry}(c)
is obtained using the finite element method, implemented in CST Studio
Suite by DS SIMULIA. The dispersion diagram is constructed by extracting
the S-parameters of a four-port unit cell and calculating the eigenmode
using the associated transfer matrix based on the method discussed
in \cite{abdelshafy_exceptional_2019}. The results show that various
DBEs occur, at frequencies of 2.75 GHz and 3.02 GHz at $kd=\pi$,
and 5.33 GHz and 5.59 GHz at $kd=0$, where all four coalescing wavenumbers
are \textit{almost} real and equal to each other. Losses prevent the
realization of a mathematically perfect DBE \cite{abdelshafy_exceptional_2019},
which is evident from the non-vanishing imaginary part of the wavenumbers
at the DBE frequencies in Fig. \ref{fig:Example-geometry}(c). However
the main feature of the four coalescing eigenvectors is still retained
as discussed in \cite{abdelshafy_exceptional_2019} using the concept
of the eigenvector coalescing parameter (also called hyperdistance).

\begin{figure}[t]
\centering{}\centering\includegraphics[width=0.85\columnwidth]{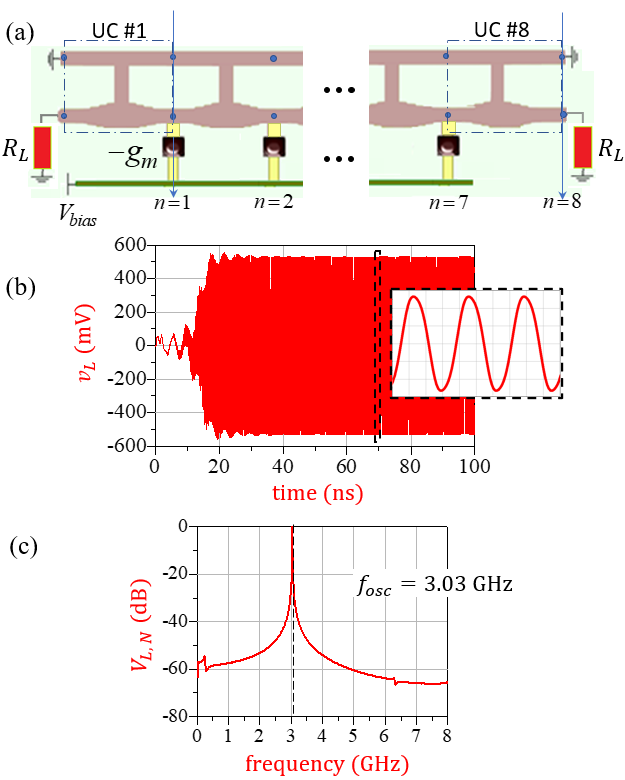}\caption{\label{fig:3GHz_oscillation} (a) Loaded DBE oscillator consisting
of 8 cascading unit cells (UCs) of microstrip-based CTLs shown in
Fig.1(b). Active devices are placed between each two adjacent unit
cells to the bias line. Two loads of $50\,\Omega$ are attached at
the two ends of the lower TL while the upper TL is terminated in short
circuits. The oscillation starts for sufficiently large $g_{m}=3\,$mS.
(b) Voltage waveform $v_{L}(t)$ monitored at a $50\,\Omega$ load
where steady state oscillation is observed in less than 30 ns. (c)
The normalized spectrum $V_{L,N}(f)$ shows that oscillations occur
at 3.03 GHz, that is very close to the DBE frequency of 3.02 GHz in
Fig. 1(c). The spectrum is calculated by applying the Fourier transform
in a time window from 35 to 100 nsec.}
\end{figure}

The distributed DBE oscillator is realized by incorporating discrete
active components in a cavity made of a finite-length CTL exhibiting
a DBE. Each active component is arranged between two adjacent unit
cells to balance small distributed losses and compensate for the two
load terminations, and hence start the oscillation as shown in Fig.
\ref{fig:3GHz_oscillation}(a). The high quality factor of such DBE
cavities and the concept of DBE resonance has been already explained
in \cite{othman_low_2016,nada_giant_2018}. Gain is modeled using
the non-linear cubic I-V characteristic $i(t)=-g_{m}v(t)+\zeta v^{3}(t)$
of the active device \cite{oshmarin_new_2019} which can be practically
implemented with circuits with amplifying devices, such as CMOS transistors
or op-amps, with positive feedback. Here $-g_{m}$ is the small-signal
slope of the I-V curve in the negative resistance region, and $\zeta$
is the third-order non-linearity constant that models the saturation
characteristic of the device. \textcolor{black}{To mimic the clipping
mechanism in realistic active devices we choose the turning point
(i.e., which is a point in the I-V curve of the active device above
which it starts to behave as a passive one; more details} \textcolor{black}{are
in \cite{oshmarin_new_2019}) $v_{b}=\sqrt{g_{m}/3\zeta}$ of the
I-V characteristics to be 1 volt, and accordingly, we set $\zeta=g_{m}/3$.
The smaller the $\zeta$ value the higher the output swing and the
time required to reach steady state.} 

\textcolor{black}{In general, active components are realized either
as a single-ended or differential topology. The most common single-ended
realizations are based on two-terminal diodes such as the Gunn diode
\cite{gunn1963microwave,ridley1961possibility}, IMPATT diode \cite{shockley1954negative},
and tunnel diode \cite{carroll1963tunnel}. Such elements are mainly
implemented in a compound semiconductor (III-V semiconductor) and
used in high power oscillators \cite{gunn1963microwave}. On the other
hand, a negative differential resistance is most commonly realized
in CMOS technology as a cross-coupled transistor pair \cite{razavi_design_2012,razavi_rf_1998}.
As long as the cross-coupled pair is biased by a tail current source,
a single-ended version can also be realized, with the other terminal
connected to an appropriate dc voltage. }

The following calculations and simulations are carried out using the
time-domain transient solver implemented in Keysight Advanced Design
System (ADS) software by means of scattering parameters obtained through
full-wave simulation, where the excitation is modeled by thermal noise
generation in the load resistors. The schematic of the proposed distributed
oscillator is shown in Fig. \ref{fig:3GHz_oscillation}(a) where the
active devices are attached to the lower transmission line in between
adjacent unit cells toward the ground (the bias line). For simplicity,
we assumed that the gain $-g_{m}$ is equal in all the active devices.
In fact, the overall performance of this kind of distributed oscillators
can be improved by optimizing the distribution of the gain values
of active devices along the finite structure.

To determine the oscillation threshold, which is the minimum value
of the gain conductance $g_{m}$ to start oscillations, we tested
the finite-length loaded cavity shown in Fig. \ref{fig:3GHz_oscillation}(a)
with both ends of the lower TL terminated with $50\,\Omega$ and both
ends of the upper TL terminated by short circuits to ground. The oscillation
threshold is obtained by sweeping the gain $-g_{m}$ value until the
oscillation starts. Accordingly, we report that the oscillation threshold
is $g_{m,th}=1\,$mS for the 8 unit cells oscillator in Fig. \ref{fig:3GHz_oscillation}(a).
Note that, the oscillation threshold value depends on the length of
the finite structure (i.e. number of unit cells) and the load values
$R_{L}$. Therefore, since we analyze the effect of load variation,
for the rest of the paper we choose $g_{m}=3\,\textrm{mS}$ to be
sufficiently larger than the oscillation threshold $g_{m,th}$ for
a large value of $R_{L}=1\textrm{\,M}\Omega$. The waveform $v_{L}(t)$
at either load reaches a steady state in less than 30 ns as shown
in Fig. \ref{fig:3GHz_oscillation}(b). The oscillation frequency
is determined by Fourier transforming $v_{L}(t)$ in the time window
from 35 to 100 ns, shown in Fig. \ref{fig:3GHz_oscillation}(c), and
it clearly confirms the single-frequency oscillatory behavior despite
the length of the cavity and the presence of seven active devices.

Fig. \ref{fig:3 V,I  distribution} shows the magnitude of voltage
and current distributions in the loaded CTL cavity of Fig. \ref{fig:3GHz_oscillation}(a).
These voltages and currents are evaluated at nodes $n=1,2,\ldots,8$,
in the presence of the active devices with $g_{m}=3\,$mS. It can
be observed that the voltage reaches its peak magnitude in the middle
of the cavity; the voltage magnitudes in the lower TL are approximately
four times larger than those in the upper TL.

\begin{figure}[t]
\centering{}\centering\includegraphics[width=0.75\columnwidth]{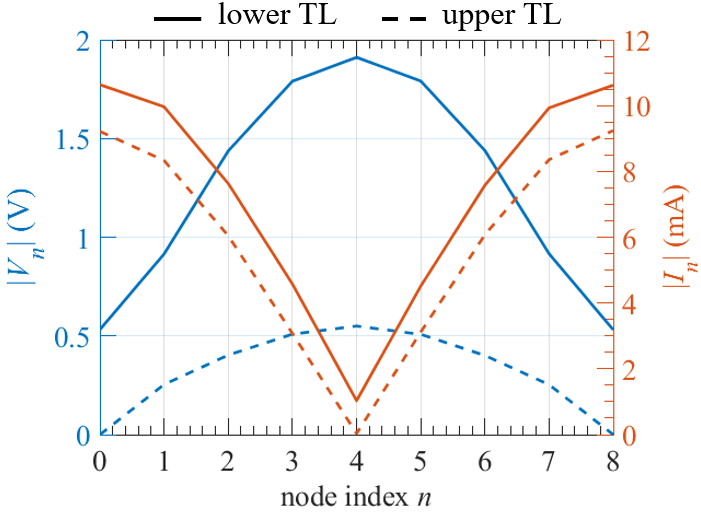}\caption{\label{fig:3 V,I  distribution} (a) Voltage and current distributions
on top and bottom TLs of the distributed DBE oscillator shown in Fig.
2(a), when the bottom TL is terminated in two $50\,\Omega$ loads,
while the top TL is short circuited at its two ends. ($|V_{n}|$ and
$|I_{n}|$ represent voltage and current time-harmonic waveform peak
values, which are phasors' amplitudes, at steady state regime).}
\end{figure}

\begin{figure}[t]
\centering{}\centering\includegraphics[width=0.85\columnwidth]{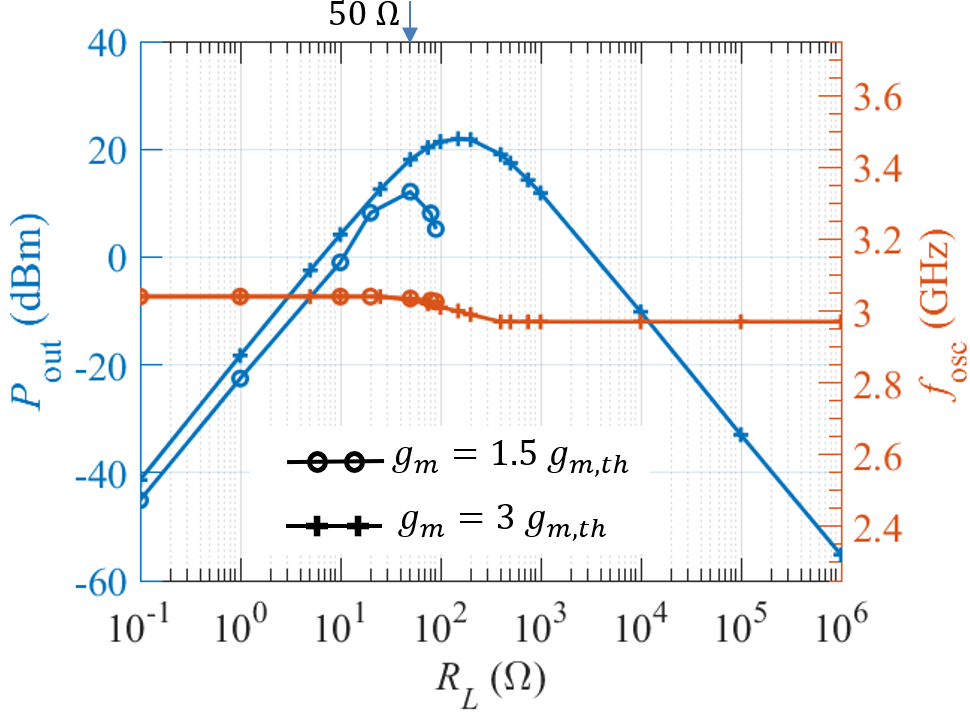}\caption{\label{fig:4 o/p power and oscillaton frequency versus load} Average
output power and oscillation frequency versus load resistance for
the distributed DBE oscillator shown in Fig. 2(a), for $g_{m}=3\,$mS
and $g_{m}=1.5\,$mS. The stability of the oscillation frequency over
a huge variation of the load resistance shows an important advantage
of the proposed oscillator: the frequency of oscillation is almost
the same, i.e. \textasciitilde 3 GHz, with a very slight shift that
does not exceed $\pm1$\% (\textasciitilde 30 MHz).}
\end{figure}

An important advantage of the proposed DBE oscillator is the robustness
of the oscillation frequency against a large variation in the load.
\textcolor{black}{This robustness is directly related to the strong
DBE resonance associated to the cavity made of a finite-length CTL
exhibiting a DBE \cite{figotin_gigantic_2005,othman_giant_2016,oshmarin_new_2019}.
}This advantage has been shown only for a DBE-based double ladder
lumped-element circuit oscillator with only one active device \cite{oshmarin_new_2019}.
Typically, the oscillation behavior is very sensitive to the output
termination resistance variation resulting in a significant shift
in oscillation frequency (e.g. mode jumping in ladder oscillators
\cite{endo_mode_1976,sloan_theory_2017}); in some cases the oscillation
stops. Fig. \ref{fig:4 o/p power and oscillaton frequency versus load}
shows the effect of varying the load resistance on the oscillation
frequency and on the average output power in the proposed distributed
DBE oscillator, for $g_{m}=3\,$mS. The result shows a stable frequency
of oscillation with a change of only 1\% over a change of load resistance
over 7 orders of magnitude. The same plot also shows the total output
power on both loads as a function of the load resistance, where the
maximum output power corresponds to $R_{L}=150\,\Omega$. Note that,
the distributed DBE oscillator also shows a stable frequency of oscillation
when changing the gain as long as it exceeds the threshold to start
the oscillation. This is shown by plotting again the output power
and frequency of oscillation for the smaller conductance of $g_{m}=1.5\,$mS.
For $g_{m}=1.5\,$mS, the maximum output power occurs when $R_{L}=50\,\Omega$
and the oscillation stops when the loads $R_{L}\geq100\,\Omega$ (as
$1.5\textrm{\,mS}<g_{m,th}|_{R_{L}=100\,\Omega}$). \textcolor{black}{Moreover,
the robustness of the oscillation frequency against a wide variation
in the gain is shown in Fig. 5(a), by fixing the two load resistances
to $R_{L}=50\,\Omega$ and varying the gain conductance from $g_{m,th}<g_{m}<9\,g_{m,th}$.
It is clear from the red curve in Fig. 5(a) that increasing the gain
in the proposed oscillator has a negligible effect on the oscillation
frequency (i.e. less than 1\% over the studied range while for $g_{m}>9\,g_{m,th}$
the oscillation frequency jumps to a higher frequency at 7.6 GHz),
whereas the output power (blue curve) increases by increasing the
gain conductance.} The stability analysis was repeated using a single
load resistance, with the other end short circuited, which leads to
very similar results and an even high stability of the oscillation
frequency over load variation (not reported here for the sake of brevity).

\begin{figure}[t]
\centering{}\centering\includegraphics[width=0.8\columnwidth]{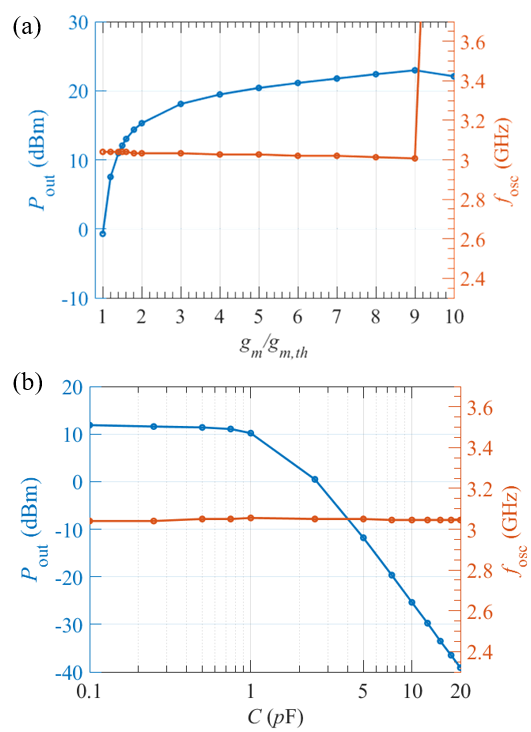}\caption{\label{fig:5 o/p power and oscillaton frequency versus load-1} \textcolor{black}{Average
output power and oscillation frequency versus (a) gain variation,
and (b) capacitive loads variation. In (b) the capacitor $C$ is parallel
to each of the two load resistors $R_{L}=50\,\Omega,$assuming $g_{m}=1.5\,$mS.}}
\end{figure}

\textcolor{black}{To better understand the robustness of the oscillator
to load variations, we add a capacitance $C$ in parallel to each
of the two load resistors $R_{L}=50\,\Omega$ shown in Fig. \ref{fig:3GHz_oscillation}(a)
to imitate the effect of a parasitic capacitance. Fig. \ref{fig:5 o/p power and oscillaton frequency versus load-1}(b)
shows the effect of varying such load capacitance on the oscillation
frequency and on the average output power, for $g_{m}=1.5\,g_{m,th}=1.5\,\text{mS}$.
The variation of the oscillations frequency is negligible for all
shown $C$ values.}

\section{Conclusion}

It has been shown that the DBE in coupled periodic waveguides is useful
to conceive new schemes for arrays of coupled oscillators. The DBE
structure in the cavity made of a periodic waveguide strongly synchronizes
a discrete set of oscillators resulting in an overall single mode
of oscillation. The single oscillation frequency, in close proximity
to the DBE frequency, has been theoretically demonstrated through
full-wave transient simulations. Results demonstrate the stability
of the oscillation frequency over a very wide range of load or gain
variation, to confirm the stable single-frequency oscillation scheme
dictated by the modal degeneracy.

This new scheme of operation is promising for boosting the overall
performance of RF sources, where potential benefits include spectral
purity and high power efficiency. This scheme based on the DBE concept
would be valuable in devices that require several coherent sources,
in power combining, with a possible extension to distributed radiating
active antenna arrays at microwaves and millimeter waves.

\bibliographystyle{IEEEtran}

\bibliography{MyLibrary}

\end{document}